\documentclass[12pt]{article}
\usepackage{graphicx,cite,epsfig}
\def\baselinestretch{1.5}
\newcommand{\ba}{\begin{array}}
\newcommand{\ea}{\end{array}}
\newcommand{\bd}{\begin{displaymath}}
\newcommand{\ed}{\end{displaymath}}
\newcommand{\be}{\begin{equation}}
\newcommand{\ee}{\end{equation}}
\def\bt{\begin{table}}
\def\et{\end{table}}
\def\bi{\begin{itemize}}
\def\ei{\end{itemize}}
\def\bea{\begin{eqnarray}}
\def\eea{\end{eqnarray}}

\def\l{\lambda}

\def\N0{\widetilde{\chi}^0}

\def\Cpm{\widetilde{\chi}^\pm}

\def\a{\alpha}
\def\as {\alpha_s}
\def\b{\beta}
\def\g{\gamma}
\def\d{\delta}

\def\m{\mu}

\def\G{\Gamma}

\def\s{\sigma}

\def\q2 {q^2}

\def\r {\rightarrow}

\catcode`@=11 % This allows us to modify PLAIN macros.
\def \gsim{\mathrel{\mathpalette\@versim>}}
\def \lsim{\mathrel{\mathpalette\@versim<}}
\def \@versim#1#2{\lower0.4ex\vbox{\baselineskip\z@skip\lineskip\z@skip
     \lineskiplimit\z@\ialign{$\m@th#1\hfil##\hfil$%
     \crcr#2\crcr\sim\crcr}}}
\catcode`@=12 % at signs are no longer letters
\begin{document}
\setcounter{page}{0}
\thispagestyle{empty}

% --------------------------------------------------------------------------
%                          Title And Abstract Page
%--------------------------------------------------------------------------
\begin{flushright}
HRI-P-05-10-001\\
\end{flushright}
\begin{center}
{\Large\sc Distinguishing split supersymmetry in Higgs signals at the 
Large Hadron Collider }\\[20mm]
Sudhir Kumar Gupta\footnote{E-mail: guptask@mri.ernet.in}, 
Biswarup Mukhopadhyaya\footnote{E-mail: biswarup@mri.ernet.in} and
Santosh Kumar Rai\footnote{E-mail: skrai@mri.ernet.in}\\
{\em Harish-Chandra Research Institute,\\
Chhatnag Road, Jhusi, Allahabad - 211 019, India}
\end{center}
 \vskip 20 mm

\begin{abstract} 
We examine the possibility of detecting signals of
split supersymmetry in the loop-induced decay $h\longrightarrow
\gamma\gamma$ of the Higgs boson at the Large Hadron Collider, where
charginos, as surviving light fermions of the supersymmetric spectrum,
can contribute in the loop. We perform a detailed study of
uncertainties in various parameters involved in the analysis, and thus
the net uncertainty in the standard model prediction of the rate. After
a thorough scan of the parameter space, taking all constraints into
account, we conclude that it will be very difficult to infer about
split supersymmetry from Higgs signals alone. 
\end{abstract}

%\vskip 0.5 true cm

%\newpage
\setcounter{footnote}{0}
\def\baselinestretch{1.8}
%\begin{quotation} \noindent \sl
%\end{quotation} \rm\normalsize
\vfill

\newpage
% --------------------------------------------------------------------------
%                       Body of the paper
%--------------------------------------------------------------------------
\section{Introduction}
%----------------------------------------------------------------------------
The idea of a supersymmetric nature, with supersymmetry (SUSY)  broken
in a phenomenologically consistent manner, is several decades old now.
It is expected that the Large Hadron Collider (LHC)  will reveal its
trace if the scale of SUSY breaking is within a TeV or so.  In
addition, signals for the Higgs boson(s) at the LHC are also likely to
yield useful information about SUSY. For example, in the minimal SUSY
standard model (MSSM) and most of its extensions, the lightest neutral
Higgs has a mass within about 135 GeV. Furthermore, its couplings with
the standard model particles differ from the standard model values, and
such departure can be tested at the LHC and more precisely at a linear
collider, giving us an indication about a supersymmetric world from the
Higgs signals themselves.

The situation is somewhat different in split SUSY, a recently proposed
scenario where all supersymmetric scalars are very heavy while the
gauginos and Higgsinos can be within the TeV scale
\cite{nima1,romanino}. Such a scenario is motivated by the fact that an
inadmissibly large cosmological constant is difficult to avoid in a
broken SUSY model, unless one fine-tunes parameters to a high degree.
Therefore, it has been argued, it may not be out of place to stabilize
the electroweak (EW) scale, too, via fine tuning. Nonetheless, SUSY as 
an artifact of theories such as superstring may still be around, albeit 
with a large breaking scale.

Since it does not claim to solve the hierarchy problem, split SUSY can
have the scalar masses (and the SUSY breaking scale) as high as
$10^{13}$ GeV or so. It avoids the problems with flavor changing
neutral current plaguing the usual SUSY models, still provides a dark
matter candidate, and even offers to retain gauge coupling through
TeV-scale thresholds consisting of incomplete representations of the
Grand Unification group \cite{romanino}. Thus, although the very
philosophy underlying split SUSY may be questioned, it is important to
explore its observable consequences. In particular, one would always
like to see if the Higgs sector still contains information on new
physics. The problem is that in the split SUSY scenario, the low-energy
spectrum contains only one neutral Higgs, its interaction strength with
all standard model particles being exactly as in the standard model
itself. This makes it difficult to distinguish split SUSY from signals
of the Higgs boson, at least in tree-level processes, since such
processes are unlikely to produce SUSY particles from decays of the
Higgs.

It has been suggested earlier \cite{spshiggs} that it may be possible
to recognize a Higgs in such a case through its loop-induced decays.  
In particular, the decay channel $h\longrightarrow \gamma\gamma$ gets
additional contributions from chargino loops. If these contributions
are substantial, then it may be possible, it has been argued, to seek
the signature of split SUSY in the two-photon decay channel of the
Higgs boson, even before the accessible part of the SUSY spectrum
reveals itself.

However, the difference made by charginos in the loop-induced effects
needs to be analyzed with the full process of Higgs production and its
subsequent decay in mind. The authors themselves noted in passing in
reference \cite{spshiggs}, that the error in measurements might be
substantial at the LHC. Nonetheless, it requires a thorough analysis of
the various parameters involved, in order to ascertain whether split
SUSY could leave its mark on Higgs decay in the most energetic high
energy collider approved till now. In this paper we carry out such an
analysis, taking into account all uncertainties in experimental
measurements as well as theoretical predictions. Thereafter we make a
thorough scan of the split SUSY parameter space, looking for regions
where the chargino contributions in the loop could stand out against
other uncertainties in the observed event rates. Our conclusion is that
it may be difficult to be sure of any split SUSY contributions over
most of the parameter space of one's interest.

In section 2 we outline the relevant features of split SUSY. In section
3 we take up signals for the Higgs boson, where the diphoton decay mode
and the relevant procedure for predicting it are discussed. The various
uncertainties in the standard model prediction, relevant for our study,
are listed in section 4, while section 5 contains the results of a
numerical scan of the parameter space. We summarise and conclude in
section 6.

%--------------------------------------------------------------------------
\section{The split SUSY spectrum}
As has been mentioned in the previous section, this scenario introduces
a splitting between scalars and the fermions. This means, all the
squarks and sleptons as well as all physical states in the electroweak
symmetry breaking sector excepting one are ultra-heavy, while gauginos,
Higgsinos and one (finely-tuned) neutral Higgs boson remain light.

%========================================
The low-energy spectrum of split SUSY can be obtained by writing the
most general renormalizable Lagrangian \cite{romanino} where the heavy
scalars have been integrated out and only one Higgs doublet $(H)$ is
retained:
\bea
{\cal L}&=&m^2 H^\dagger H-\frac{\lambda}{2}\left( H^\dagger H\right)^2
-\left[ h^u_{ij} {\bar q}_j u_i\epsilon H^* +h^d_{ij} {\bar q}_j d_iH
+h^e_{ij} {\bar \ell}_j e_iH \right. \nonumber \\
&&+\frac{M_3}{2} {\tilde g}^A {\tilde g}^A +\frac{M_2}{2} 
{\tilde W}^a {\tilde W}^a +\frac{M_1}{2} {\tilde B} {\tilde B}
+\mu {\tilde H}_u^T\epsilon {\tilde H}_d \nonumber \\
&&\left. +\frac{H^\dagger}{\sqrt{2}}\left({\tilde g}_u \sigma^a {\tilde W}^a 
+{\tilde g}_u^\prime {\tilde B} \right) {\tilde H}_u
+\frac{H^T\epsilon}{\sqrt{2}}\left(
-{\tilde g}_d \sigma^a {\tilde W}^a
+{\tilde g}_d^\prime {\tilde B} \right) {\tilde H}_d +{\rm h.c.}\right] ,
\label{lagr}
\eea
where $\epsilon =i\sigma_2$ and ${\tilde H}_{u,d}$ (Higgsinos), $\tilde
g$ (gluino), $\tilde W$ (W-ino), $\tilde B$ (B-ino) are the gauginos.

The coupling strengths of the effective theory at the scale ${m_S}$,
where ${m_S}$ is the scale of SUSY breaking, are 
obtained by matching the Lagrangian in equation~\ref{lagr}. with the 
interaction terms of the supersymmetric Higgs doublets $H_u$ and $H_d$:
\bea
{\cal L}_{\rm susy}&=&
-\frac{g^2}{8}\left( H_u^\dagger \sigma^a H_u + H_d^\dagger \sigma^a
H_d \right)^2
-\frac{g^{\prime 2}}{8}\left( H_u^\dagger H_u - H_d^\dagger  H_d
\right)^2 \nonumber \\
&&+Y^u_{ij}H_u^T\epsilon {\bar u}_i q_j
-Y^d_{ij}H_d^T\epsilon {\bar d}_i q_j
-Y^e_{ij}H_e^T\epsilon {\bar e}_i \ell_j
\nonumber \\
&&-\frac{H_u^\dagger}{\sqrt{2}}\left( g \sigma^a {\tilde W}^a
+g^\prime {\tilde B} \right) {\tilde H}_u
-\frac{H_d^\dagger}{\sqrt{2}}\left(
g \sigma^a {\tilde W}^a
-g^\prime {\tilde B} \right) {\tilde H}_d +{\rm h.c.}
\label{lagrs}
\eea
The combination  $H=-\cos\beta \epsilon H_d^*+\sin\beta H_u$
is then fine-tuned to have a small mass term. The matching conditions
for the coupling constants in equation~\ref{lagr}. at the scale ${m_S}$ are
obtained by replacing $H_u\to \sin\beta H$, $H_d\to \cos\beta \epsilon
H^*$ in equation~\ref{lagrs}.:
\bea
\lambda(m_S)& =& \frac{\left[ g^2(m_S)+g^{\prime 2}(m_S)        
\right]}{4} \cos^22\beta,
\label{condh}\\
h^u_{ij}(m_S)= Y^{u*}_{ij}(m_S)\sin\beta , &&
h^{d,e}_{ij}(m_S)= Y^{d,e*}_{ij}(m_S)\cos\beta ,\\
{\tilde g}_u (m_S)= g (m_S)\sin\beta ,&&
{\tilde g}_d (m_S)= g (m_S)\cos\beta ,\\
{\tilde g}_u^\prime (m_S)= g^\prime (m_S) \sin\beta ,&&
{\tilde g}_d^\prime (m_S)= g^\prime (m_S)\cos\beta ,
\label{condg}
\eea
%========================================
\noindent
where $\lambda$ is the scalar self-coupling of a theory with a single
Higgs doublet, $g$, $g^\prime$ are gauge couplings, and $Y$'s are the
Yukawa couplings of the two doublets at the scale $m_S$. The Yukawa
interactions of the surviving Higgs doublet below $m_S$ is obtained
from the matching conditions and are denoted by $h^{(u,d,e)}$.

The low energy effective Lagrangian, as already stated, contains only
the neutral CP-even Higgs, a physical state which is henceforth denoted
by $h$. Its relevant coupling is obtained by setting $\b - \a~ =~ \pi
/2$ in the two-Higgs doublet Lagrangian, which is equivalent to the
decoupling limit. Gauge and Yukawa couplings at low energy are exactly
as in the standard model, though these can be obtained from the
original Lagrangian in the said limit, through evolution from the scale
$m_S$ using the matching conditions mentioned before.

Similarly, the Higgs mass at EW scale is governed by 
the quartic coupling and the vev $v$:
\bea
m_h \sim \sqrt{\lambda} v    
\eea 
where the low-energy Higgs quartic coupling is controlled by the
logarithmically enhanced contribution given by the evolution of
$\lambda$ from the high scale $m_S$, for which the boundary value is
given by equation~\ref{condh}. In this scenario, one can make the Higgs
heavier than the lightest neutral supersymmetric Higgs boson
\cite{nima1,romanino,sphmass}.  Thus, by taking the maximum value of
$m_S$ to be about $10^{13}$ GeV (for which the justification is given
below), it is possible to have a Higgs of mass upto about 170 GeV
\cite{sphmass}, making the scenario phenomenologically less restrictive
from the viewpoint of Higgs searches.

Theoretically, the fermions can be visualized as being protected by an
R-symmetry or a Peccei-Quinn symmetry \cite{nima1,romanino}. In order
to make one physical Higgs state light, one has to fine-tune in the
Higgsino mass parameter $\mu$, the bilinear soft parameter $\mu B$ and
the two soft SUSY breaking mass terms for the two doublets, although
the viability of such tuning has sometimes been questioned
\cite{drees}. In general, a number of theoretical proposals have been
made concerning the origin of a split spectrum and some of its
consequences \cite{nima2,spspectrum}.

A number of phenomenological consequences of a split spectrum have been
studied in the literature \cite{sppheno,spgluino,sprare,spfermion}.  
For example, gluinos can be long-lived since their decays are mediated
by the squarks whose masses are at the SUSY breaking scale. The
collider implications of such long-lived gluinos as well as of heavy
sleptons vis-a-vis light charginos and neutralinos have been already
reported \cite{spgluino,sprare,spfermion}. Also, an upper limit of
about $10^{13}$ GeV on the SUSY breaking scale has been suggested from
the consideration that gluino lifetime has to be shorter than the age
of the universe \cite{nima1}.  Also, various constraints on the
scenario ensue from potentially long-lived `R-hadrons' containing
gluinos in a split SUSY scenario \cite{nima1}. In models based on
supergravity, implications on the gravitino mass and dark matter have
been discussed as well \cite{darkmatter}. The possible enhancement of
fermion electric dipole moments has also been reported
 \cite{spedm}. In addition, it has been seen that R-parity violation
in split SUSY can lead to extremely interesting situations where either
the lightest neutralino can still be a dark matter candidate through its
long lifetime, or it can appear invisible in collider experiments while
not contributing to the relic density of the universe \cite{gkm}.

In addition to the gaugino and Higgsino mass parameters,
the trilinear soft-breaking term $A$ etc. which are all within a TeV, the  
split SUSY spectrum depends on the SUSY breaking scale, in the sense 
that boundary conditions for parameters affecting low-energy physics
are set at that scale. For example, the quantity $\tan\beta$ can no more
be interpreted as the ratio of vacuum expectation values (vev) of the two 
scalar doublets, simply because one of the doublets is integrated out 
when electroweak symmetry breaking takes place. It is instead more
sensible to treat the angle $\beta$ as a parameter specifying the linear
combination of the two doublets that survive till the EW scale.
The relevant parameters (such as $\tilde{g_u}/g,\tilde{g_d}/g$ etc.)
which enter the chargino mass matrix at low-energy are obtained via 
evolution from the scale $m_S$ (where they are related to the angle
$\beta$). This evolution has to be taken into account whenever a reference 
to physics at the scale $m_S$ has to be made. 

\section{Higgs signals and the diphoton mode}

If the Higgs exists in the mass range expected in split SUSY, we 
will be able to  see it during an early phase of the LHC. 
The question that arises is whether it can be distinguished from
the standard model Higgs. If that is possible, then it will be 
an indication of new physics in Higgs signal itself, even if 
the detection of the new particles in the spectrum are delayed,
due, for example, to their high mass.

As we have seen above, all tree-level interactions revealing the Higgs
at the LHC are exactly as in the standard model. Therefore, we must
examine loop induced Higgs decay processes where virtual SUSY particles
may contribute. The most suggestive channel in this context is the
standard production of the Higgs followed by its decay into the
diphotons. In this mode, the (partial) decay width $\Gamma(h \r \g
\g)$, gets additional contributions from chargino loops. Recently, it
has been suggested \cite{spshiggs} that in some regions of the
parameter space these loop contributions may alter the Higgs decay
widths by a few per cents, thus making it distinguishable from the
standard model Higgs boson.

It has to be remembered, however, that the above decay width is not a
directly measurable quantity at the LHC. This is because the width is
of the order of keV in the relevant Higgs mass range, which is smaller
than the resolution of the electromagnetic calorimeters to be used
\cite{atlas,cms}. Therefore, it is not clear {\it prima facie} how well
the signature of split SUSY can be extracted in this channel, given the
rather sizable theoretical as well as experimental uncertainties in the
various relevant parameters.

We, therefore, have chosen to do a calculation involving the full
process $(p p \r h X \r \g \g)$, that is to say, the production of the
Higgs followed by its decay into the diphoton final state. Taking all
uncertainties into account, we have tried to find the significance
level at which the chargino-induced contributions can be differentiated
in different regions of the parameter space. We have confined ourselves
to the production of Higgs via gluon fusion. The other important
channel, namely gauge boson fusion, has been left out of this study,
partly because it is plagued with uncertainties arising, for example,
from diffractive production, which may be too large for the small
effects under consideration here.

In the standard model, the decay rate of the Higgs boson to a photon
pair is driven by loop-induced contributions from all charged particles
as shown in figure 1.  Dominant among them are the loops driven by the
W boson and the top quark, although contributions from the bottom and
charm quarks as well as the $\tau$-lepton cannot be ignored in a
precision analysis. The contributions from such loops, including QCD as
well as further electroweak corrections, are well-documented in the
literature \cite{qcdcorr,2lewc}.

The additional contributions from charginos depend on interactions that
can be extracted from the split SUSY effective Lagrangian:
\be 
L \supset - \frac{H^\dagger}{\sqrt 2}({\tilde g_u} \s^a {\tilde W^a}
+{\tilde g_u^\prime} {\tilde B}) {{\tilde H}_u}
-\frac{H^T}{\sqrt 2} {i \s_2}(-{\tilde g_d} \s^a {\tilde W}^a
+{{\tilde g_d}^\prime} {\tilde B}) {\tilde H_d} + h.c.
\ee
Representative Feynman graph relevant for the process is shown in 
figure 2.  

Using the above Lagrangian, one obtains the following 
contribution to the above decay rate, as a sum of the standard and
chargino-induced diagrams:
%%%%%%
\bea \Gamma(h \to \gamma\gamma) = \frac{G_F}{128\sqrt{2}}
\frac{\alpha^2 m_h^3}{\pi^3}\left|\sum_{i} A_i\right|^2 \eea
%%%%%%
where $i$ stands for different particles in the loop. The 
amplitudes $A_i$ are
%%%%%%
\bea
 A_W &=& C_W F_1(\l_W)    \nonumber \\ 
 A_f &=& N_c^f Q_f^2 C_f F_{1/2}(\l_f)  \nonumber \\ 
 A_{\Cpm} &=& C_{\Cpm} \frac{m_W}{m_{\Cpm}} F_{1/2}(\l_{\Cpm})
\label{amp}
\eea
%%%%%%
where $\l_i = \frac{4 m_i^2}{m_h^2}$, $m_i$ being the mass of the
particle inside the loop. The functions $F_1,~F_{1/2}$ are
given by
%%%%%%
\bea
F_1(\l) &=&  3\l + 2 + 3\l(2 - \l)f(\l) \nonumber \\
F_{1/2}(\l)&=& -2\l\left[1 + (1 -\l)f(\l)\right] 
\eea
%%%%%%
The function $f(\l)$ depends on the value of $\l$ and takes the form:
\bea
f(\l) = \left[\sin^{-1}\sqrt{\frac{1}{\l}}\right]^2 ~~{\rm for}~~\l\ge 1
\nonumber \\
f(\l) = -\frac{1}{4}\left[\log\left(\frac{1+\sqrt{1-\l}}{1-\sqrt{1-\l}}\right) -
i\pi \right]^2 ~~ {\rm for}~~\l < 1
\label{absorptive}
\eea
\begin{figure}
\begin{center}
\epsfig{file=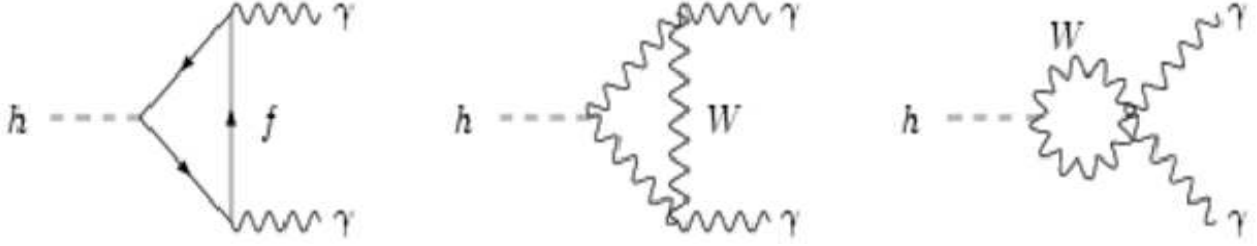
,width=7.5in,height=1.3in}
\caption{\it{Standard model Feynman graphs contributing to 
the process $h\to \gamma\gamma$ at the lowest order.}}
\label{smfg}
\end{center}
\end{figure}
\begin{figure}
\begin{center}
\epsfig{file=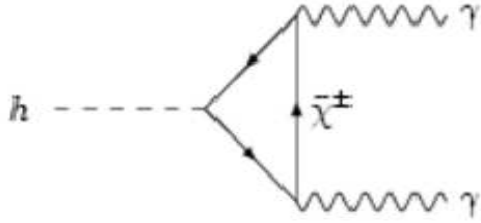
,width=7.5in,height=1.3in}
\caption{\it{Additional contribution to 
$h\to \gamma \gamma$ due to the chargino loops in split SUSY.}}
\label{spsfg}
\end{center}
\end{figure}
The colour factor $N_c^f$ equals 3 for quarks and 1
for leptons. One has $C_W = C_f = 1$, while the chargino coupling is
given by,
$$C_{\Cpm} = 2~( S_{ii}~\tilde{g_u}/g + Q_{ii}~\tilde{g_d}/g)$$
with $S_{ij} = U_{i1}V_{j2}/\sqrt{2}$ and $Q_{ij} = U_{i2}V_{j1}/\sqrt{2}$.
The matrices $\bf{U}$ and $\bf{V}$ diagonalize the chargino mass
matrix. In our case $i=1$ and 2 yield the two physical charginos in the 
loops.

One has to remember that the seed parameters corresponding to split
SUSY and MSSM are fixed at the SUSY breaking scale $m_S$ and that those
featuring in the above expressions are the results of evolution down to
the EW scale ($m_W$) through renormalization group (RG)
equations \cite{nima1,romanino}. However, their low-energy values
themselves can be used in the present analysis, without any further
reference. We have also assumed gaugino unification, having the
low-energy SU(2) gaugino mass $M_2$ as an independent parameter.  Thus
the basic parameters for us are (in addition to those of the standard
model) the Higgs mass $m_h$ and the SUSY parameters $M_2$, $\mu$ (the
Higgsino mass)  and $tan\b$ . The latter, not having anything to do
with low-energy couplings of the Higgs, can easily evade the bound of
$\simeq 2$ obtained from on the measurements of Higgs mass at the Large
Electron-Positron (LEP) collider \cite{LEP2}. However, since
$\tan\beta$ governs the high-scale Lagrangian and therefore the
boundary conditions for the spectrum at the scale $m_S$, bounds of the
order of 0.5 on its value have been derived from considerations such as
the infrared fixed point for the top quark mass \cite{romanino,irfix}.
It may be noted that a similar lower bound of about 1.2 can be given on
$\tan\beta$ in the MSSM, but it is overridden by the experimental
limit. The remaining split SUSY parameters have also been restricted by
the lower bound of about 103.5 GeV on the chargino mass
\cite{chargino}.

The rate for the inclusive process $$ p p \r h ~+~ X \longrightarrow \g
\g$$ (where Higgs production takes place via gluon fusion) can be
expressed in the leading order as
\begin{equation}
R~=~ \frac{\pi^2}{8 m_h s} \frac{\G_{h \r 2g} \G_{h \r 2\g}}{\G_{tot}}\int^{1}_{\tau} {d\zeta 
\frac{1}{\zeta} g\left(\zeta,m^2_h\right)~g\left(\frac{\tau}{\zeta},m^2_h\right)}
\end{equation}
where~ $\tau =\frac{m^2_h}{S}$ and $g\left(\zeta,m^2_h\right)$ is the
gluon distribution function evaluated at $Q^2~=~m^2_h$ and parton
momentum fraction $\zeta$. $ \G_{h \r 2\g}$ and $\G_{tot}$ stand
respectively for the diphoton and total decay widths of the Higgs. The
lowest order estimate given above is further multiplied by the
appropriate K-factors to obtain the next-to-next leading order (NNLO)
predictions in QCD. While the computation of the rate is
straightforward, we realise that the various quantities used are beset
with theoretical as well as experimental uncertainties \cite{zepnew}.
We undertake an analysis of these uncertainties in the next section.

%--------------------------------------------------------------------------
\section{Numerical estimate: uncertainties} 
%--------------------------------------------------------------------------

        As has already been stated in the previous section, 
the rate for diphoton production through real Higgs at LHC is given by
\bea
R~=~ \s(p p \r h) \times B~=~\s(p p \r h)
{\frac {\Gamma(h\r\gamma\gamma)}{\Gamma_{tot}}}
\eea

We have performed a parton-level Monte Carlo calculation for the
production cross-section, using the MRS \cite{mrst} parton distribution
functions and multiplied the results with the corresponding NNLO
K-factors \cite{kfactor,hocorrs}.  It may be noted that NNLO K-factors
are not yet available for most other parameterizations. In estimating
the statistical uncertainties in the experimental value \cite{expnos},
MRS (at leading order) distributions have been used by the CMS group
while ATLAS uses CTEQ distributions.  We have obtained the aforesaid
uncertainty by taking the estimate based on MRS and multiplying the
corresponding event rate by the NNLO K-factor for MRS.  It may also be
mentioned that the difference between the NLO estimates of Higgs
production using the MRS and CTEQ parameterizations is rather small
($\lsim 2\%$), according to recent studies \cite{kfactor}.  Therefore,
it is expected that the NNLO estimate of uncertainties (where there is
scope of further evolution in any case) used by us will ultimately
converge to even better agreement with other parameterizations and will
not introduce any serious inaccuracy in our conclusions. The programme
HDECAY3.0 \cite{hdecay}, including $\mathcal{O}(\as^2)$ contributions,
has been used for Higgs decay computations .

The number of two-photon events seen is given by $\cal{L} R$ where
$\cal{L}$ is the integrated luminosity. $\cal{L}$ is expected to be
known at the LHC to within 2 \%. We include this uncertainty in our
calculation, although it has a rather small effect on our conclusions.

In order to estimate the total uncertainty in $R$, one has to first
obtain the spread in theoretically predicted value in the standard
model due to the uncertainty in the various parameters used. In
addition, however, there is an uncertainty in the experimental values;
although the actual level of this will be known only after the LHC run
begins, the anticipated statistical spread in the measured value can be
estimated through simulations.  These two uncertainties, combined in
quadrature, are indicative of the difference with central value of the
standard model prediction which is required to establish any
non-standard effect at any given confidence level. We have performed
such an exercise, taking the standard model calculation and that with
standard model + chargino contributions.
%-----------------------------------------------------------------------
\bt
\begin{center}
\begin{tabular}{|c|c|c|c|}
\hline\hline
{\bf{Parameter}}& {\bf{Central Value}} &{\bf{Present
Uncertainty}}&{\bf{LHC Uncertainty}}(projected)\\ \hline
$m_h$&$120 - 150 $&$-$&$0.2$ \\\hline 
$m_W$&$80.425$&$.034$&$.015$ \\\hline 
$m_t$&$172.7 $&$ 2.9$&$1.5$ \\\hline  
$m_b$&$4.62$&$.15$&$-$ \\\hline  
$m_c$&$1.42 $&$ .1 $&$-$ \\\hline  
$m_\tau$&$1.777 $&$.0003$&$-$\\\hline  
$\alpha_s$&$0.1187 $&$0.002$&$-$ \\\hline\hline
\end{tabular}
\caption {\it{Current and projected uncertainties (at LHC) in
the values of various parameters. All the masses are given in $GeV$.
The values are extracted from refs \cite{pdg,topmass,param}}}
\label{param.tbl}
\end{center}
\et
%-----------------------------------------------------------------------

Thus the total uncertainty in $R$ can be expressed as
\bea
{\left(\frac{\d R}{R}\right)}^2 ~=~ {\left(\frac{\d R}{R}\right)}_{\it th}^2 
~+~ {\left(\frac{\d R}{R}\right)}_{\it exp}^2
\eea
where the theoretical component can be further broken up as
\bea
{\left(\frac{\d R}{R}\right)}_{\it th}^2 ~=~ \frac{1}{R^2}\sum_{i}{\s^2_{R_i}}
\eea
where $\s_{R_i}$ stands for the spread in the prediction of R due to
uncertainty in the i$^{th}$ parameter relevant for the calculation.  
The sum runs over $m_h$, $m_W$, $m_t$, $m_b$, $m_{\tau}$ and $m_c$, in
addition to the uncertainty in the strong coupling $\alpha_s$. The
spread in the predicted value is predicted in each case by random
generation of values for each parameter (taken to vary one at a time)
within the allowed range. Thus we obtain $\frac{1}{R^2}{\s^2_{R_i}}$
corresponding to each parameter. One has to further include QCD
uncertainties arising via parameterization dependence of the parton
distribution functions (PDF) and the renormalisation scale. Although
NNLO calculation reduced such uncertainties, the net spread in the
prediction due to them could be as large as $\sim 15$ \%
\cite{kfactor,hocorrs,Belyaev,scalevarn} in the Higgs mass range $120
-150$ GeV. The levels of uncertainties in the various parameters, are
presented in Table \ref{param.tbl}.  In that table we have given the
uncertainties, wherever they are available, from recent and current
experiments like the LEP and the Tevatron. In addition, whatever
improved measurement, leading to smaller errors (in, say, $m_t$ or
$m_W$) are expected after the initial run of the LHC are also
separately incorporated in the table . We have used the estimates
corresponding to LHC wherever they are available.  In our calculation,
we have used two values of the combined uncertainty from PDF and
scale-dependence, namely, 15\% and 10\%, the latter with a view to
likely improvement using data at the LHC. This uncertainty is over and
above the uncertainty in $\alpha_s$ due to the error in measurement of
its boundary value at $m_Z$. Table 2 contains the finally predicted
values of ${\left(\frac{\d R}{R}\right)}$, for the two values of the
Higgs boson mass.

$R_{\it exp}$ includes statistical uncertainties, as estimated in
detector simulations with a luminosity of 100 $fb^{-1}$ \cite{expnos}.
As has been already mentioned, we have obtained benchmark values of
this quantity using the results for CMS presented in ref~\cite{expnos}
for MRS distributions at the lowest order, and appropriately improving
them with the NNLO K-factors available in the literature. The resulting
predictions for statistical error are $8.1\%$ for $m_{h}=130$ GeV, $8.6\%$ 
for $m_{h}=140$ GeV and $11.3\%$ for $m_{h}=150$ GeV.

%-----------------------------------------------------------------------
\bt
\begin{center}
\begin{tabular}{|c|c|c|}
\hline\hline
\multicolumn{3}{|c|}{\bf{Total Uncertainty in Standard Model Rate}}\\
\hline
{\bf{Higgs mass~(GeV)}}&{\bf{PDF + Scale Uncertainty$=15\%$}}&{\bf{PDF + Scale 
Uncertainty$=10\%$}}
\\\hline 
$130$&$18.5\%$&$14.7\%$\\
\hline  
$140$&$18.3\%$&$14.4\%$\\
\hline  
$150$&$19.4\%$&$15.8\%$\\\hline\hline  
\end{tabular}
\caption{\it{Expected total uncertainties in standard model rate at LHC. 
Entries in the second (third) column corresponds to total uncertainty 
from parton distributions and renormalization scale being equal to 
$15\%$ ($10\%$).}
\label{total.tbl}}
\end{center}
\et
%-----------------------------------------------------------------------     
   
Thus one is able to obtain the net ($1 \s$ level)  uncertainties in the
standard model. Next, the split SUSY contributions via chargino-induced
diagrams are calculated and added to the standard model amplitude. The
observable decay rate obtained therefrom is compared with that
predicted in the standard model taking the uncertainty into account at
various confidence levels. Thus one is able to decide whether the
chargino contributions to the diphoton rate are discernible from the
standard model contributions at a given confidence level for a
particular combination of split SUSY parameters.
%-----------------------------------------------------------------------
\bt
\begin{center}
\begin{tabular}{|c|c|c|}
\hline\hline
{\bf{$m_h$ ~(GeV)}}&{\bf{$tan\b$}}&{\bf{$m_S$ ~(GeV)}}\\
\hline 
$130$&$1.0$&$1\times10^5 - 1\times10^6$\\\cline{2-3}
$$&$1.5$&$0.7\times10^5 - 5\times10^5$\\\hline
$140$&$1.0$&$3\times10^7 - 6\times10^8$\\\cline{2-3}
$$&$1.5$&$2\times10^6 -4\times10^7$\\\hline
$150$&$1.2$&$7\times10^{12} - 9\times10^{12}$\\\cline{2-3}
$$&$1.5$&$0.9\times10^{12} -2\times10^{12}$\\\hline
\hline  
\end{tabular}
\caption{\it{Allowed ranges of $m_S$, 
corresponding to the three low-scale Higgs masses used here.}} 
\label{total.tan}
\end{center}
\et
%-----------------------------------------------------------------------  
The realistic estimate requires subjecting the predictions to some
experimental cuts aimed at maximizing the signal-to-background ratio as
well as focusing on kinematic regions of optimal observability. We
incorporate the effects of such cuts with the help of an efficiency
factor which, on explicit calculation in representative cases, turns
out to be approximately 50\%. The only assumptions required are that
the percentage error due to various parameters are the same for uncut
rates as those calculated with cuts, and that the standard and split
SUSY contributions suffer the same reduction due to cuts.  We have
checked that this holds true so long as the kinematic region is not
drastically curtailed by the cuts.

Before we end this section, it should be noted that the various
uncertainties quoted above are only benchmark values. The precise
levels of these uncertainties will be known after the LHC comes into
operation.

\section{Numerical estimate: discussions}

Our purpose is to see at what confidence levels one can distinguish the
split SUSY effects on $h \longrightarrow \gamma\gamma$. 
With this in view, we have presented, in figures 3 - 8, sets of contour 
plots in the $M_2$-$\m$ plane with different values of $\tan\beta$, for 
$m_h=130$ GeV, $140$ GeV and $150$ GeV. 

Since the low-scale parameters in this scenario originate in specific
boundary conditions at the SUSY breaking scale $(m_S)$, one needs to
emphasize that not all such parameters are consistent. 	In general, the 
value of $m_h$ is determined (modulo the uncertainties due to parameters 
such as $\alpha_s$ and top quark mass) once $tan\beta$ and $m_S$ are fixed. 
In this study, we are essentially interested in the low energy parameters 
which can make a difference from the standard model estimate. Therefore, 
for each $tan\beta$ used, we have found the scale $(m_S)$, so as to 
reproduce the Higgs mass used in the corresponding case. Such allowed 
ranges of $m_S$ are presented in the Table 3. In obtaining these values, 
the procedure adopted is as follows. Using a given value of $tan\beta$ as 
boundary conditions at $m_S$, and values of gauge couplings at the 
EW scale, one solves the renormalisation group equations, going through an 
iterative process till convergence is achieved. Then the quartic coupling 
$\lambda$ is evolved down to the EW scale, using $tan\beta$ as well as the 
gauge couplings at $m_S$ 
to determine its boundary value (see equation 3), whereby the Higgs mass 
$(m_h)$ is obtained. For each value of $m_h$ used in our numerical study, we 
have the value of $m_S$ which achieves that particluar $m_h$, for a given 
$tan\beta$. In this way we find that $m_h = 130 -150$ GeV is a 'reasonable' 
range, for which, with the given value of $tan\beta$, $m_S$ can be 
$\lsim 10^{13}$ GeV and at the same time not too close to the TeV scale. We 
have deliberately avoided imposing further constraints on $m_S$ in this 
phenomenological study. For $m_h =120$ GeV or less, $m_S$ tend to violate 
the aforesaid condition; therefore, we have started from $m_h = 130$ GeV.

The quantities 
${\tilde{g_u}}/{\tilde{g_d}}$ and ${\tilde{g'_u}}/{\tilde{g'_d}}$ are
both equal to $tan\b$ at the scale $m_S$, and thus their values at low
scale are obtained through running. Such values are used in the chargino 
mass matrix and Higgs-chargino coupling.

In the first three graphs, the total uncertainty arising from PDF as well
as the renormalisation scale has been taken to be 15\%. The results
where this uncertainty is 10\%, corresponding to a projected
convergence of different PDF parameterizations as well as improvement
over the current NNLO results, are shown in figures 6 - 8. The allowed 
regions represented by the contours are also subjected to the
restriction that the mass of the lighter chargino be above the current
experimental limit of 103.5 GeV.

The results in all the above cases show that the distinguishability
with the standard model effect is maximum for such values of $\mu$ and
$M_2$ which leads to the lowest possible chargino masses contributing
in the loops. For negative $\mu$, lower values of $|\mu|$ are allowed
by the above constraints; hence an asymmetry about $\mu=0$ is seen.
The dependence on $\tan\beta$ is also substantial. The maximum departure 
from the standard model contribution occurs for $\tan\beta = 1$. This is 
because the Higgs-chargino-chargino coupling is maximum when the 
charginos have equal admixture of the Wino and Higgsino components. 
When no CP - violating phase in the mixing is assumed, there is also a 
symmetry of the coupling under $\tan\beta \rightarrow \cot\beta$.

It is clear from the contours that the general level of expected
distinguishability of the split SUSY contributions is quite low. This
is primarily due to the uncertainty of ``PDF + renormalisation scale".
However, even if this uncertainty is brought down from 15\% to 10\%,
one notices that one is barely allowed a small area of the parameter
space for $\tan\beta ~\simeq~1$, where predicted effects are about
$2\sigma$; otherwise the results are even less optimistic. 
The distinguishability goes down considerably for high values of $\tan\beta$.
The other important source of uncertainty is in the b-quark mass 
(calculated at the scale $m_h$, with the boundary condition that the pole 
mass is $4.62$ GeV)  which affects the total width for $h\longrightarrow
b\bar{b}$. The results look even less optimistic if one remembers that
searches in, for example, the trilepton channel at the LHC are likely
to raise the experimental lower limit of the chargino mass, unless the
lighter chargino lies just beyond the LEP limit. Under such
circumstances, the confidence level for distinguishing the chargino
effects in the diphoton signal will be further diminished, and the
$2\sigma$ region will be obliterated in all likelihood.

\section{Summary and conclusions}

We have undertaken a thorough analysis of the split SUSY parameter
space to see if the channel $h\longrightarrow\gamma\gamma$ can allow
one to isolate the contributions from chargino-induced loops. In the
case of split SUSY, this is supposedly the only channel where the sole
surviving Higgs at the electroweak symmetry breaking scale can reveal
any difference with respect to its counterpart in the standard model.  
Although the chargino contribution has been already calculated, our
analysis, with all uncertainties duly incorporated in the production as
well as decay level, confirms that the measurable effects are very
small in all over the allowed parameter space. It is going to be very
difficult to achieve a $2\sigma$ difference with respect to the
standard model predictions, and that too for the value of $\tan\beta$
in the neighbourhood of 1. Thus it appears to us that the only way to
uncover split SUSY is to carry out an exhaustive search for the entire
superparticle spectrum at the LHC, unless some other ingenious method
can be devised to see the difference in Higgs couplings with the SUSY
fermions.

% ----------------------------------------------------------------------
{\bf{Acknowledgment:}}
\nonumber
We thank Anindya Datta, Aseshkrishna Datta, G.F. Giudice, A. Romanino, 
V. Ravindran and Sourov Roy for useful comments. 
%-------------------------------------------------------------

\vskip 5pt
%-----------------------------------------------------------------------
%                          References
%-----------------------------------------------------------------------

\newpage
%------------- Contour Plots----------------------------------
%------------15 Percent-- and M_h=130 GeV---------------------
\begin{figure}
\begin{center}
\epsfig{file=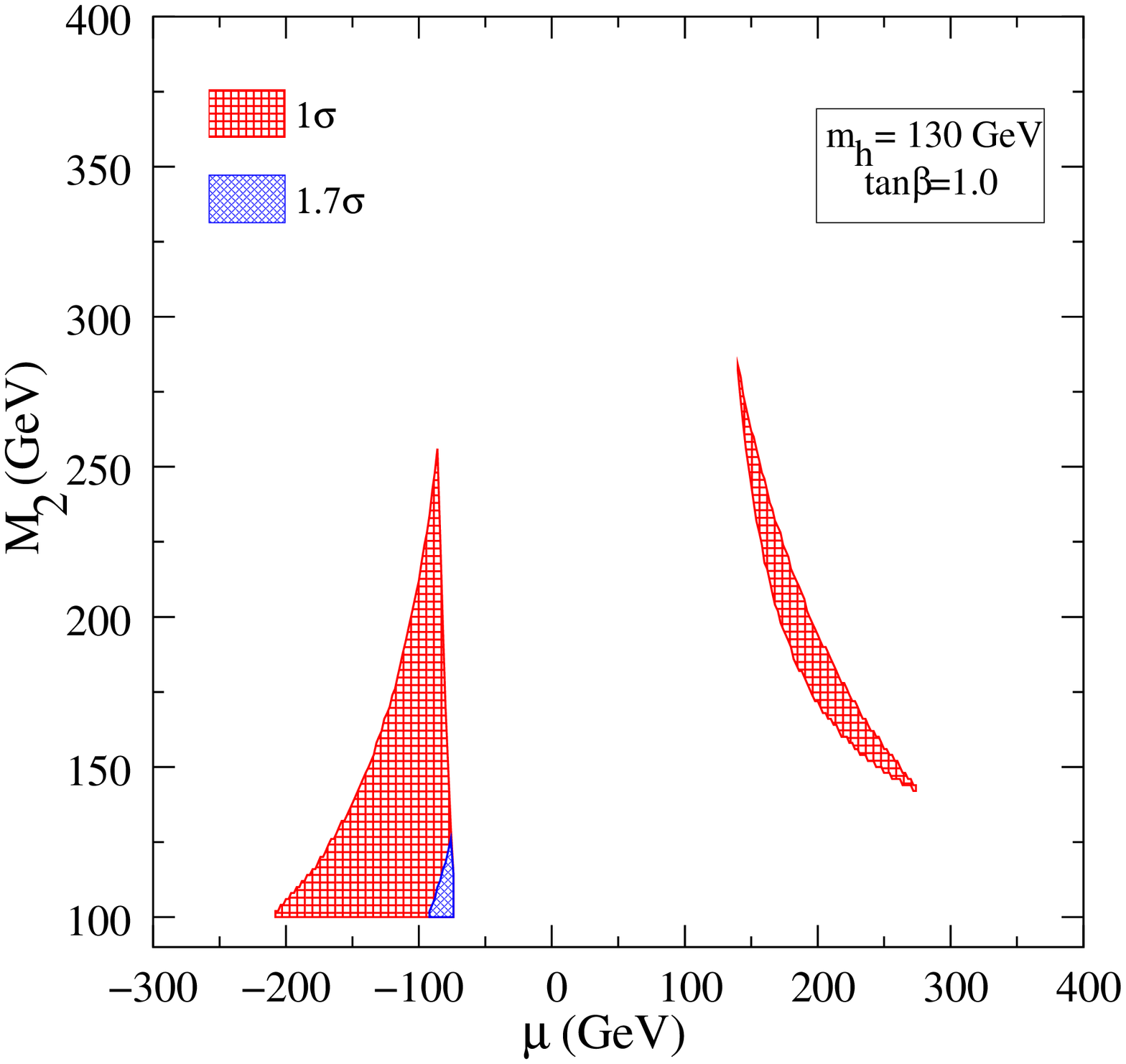,width=3in,height=3in}
\epsfig{file=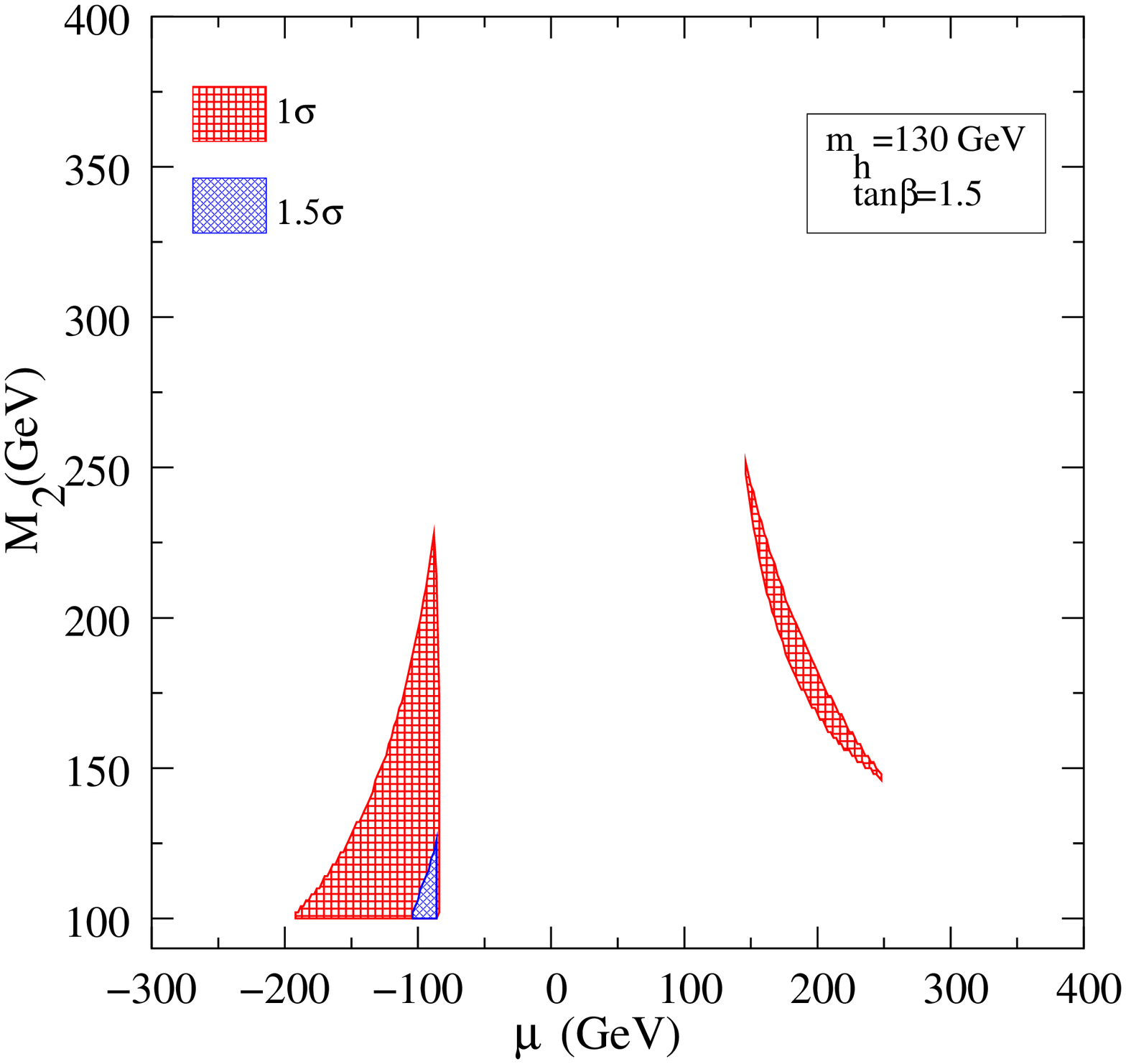,width=3in,height=3in}
\caption{\it{Contour plots for $m_h = 130$ GeV 
assuming PDF + scale uncertainty $=15\%$.}}
\label{15per130}
\end{center}
\end{figure}
%------------15 Percent-- and M_h=140 GeV---------------------
\begin{figure}
\begin{center}
\epsfig{file=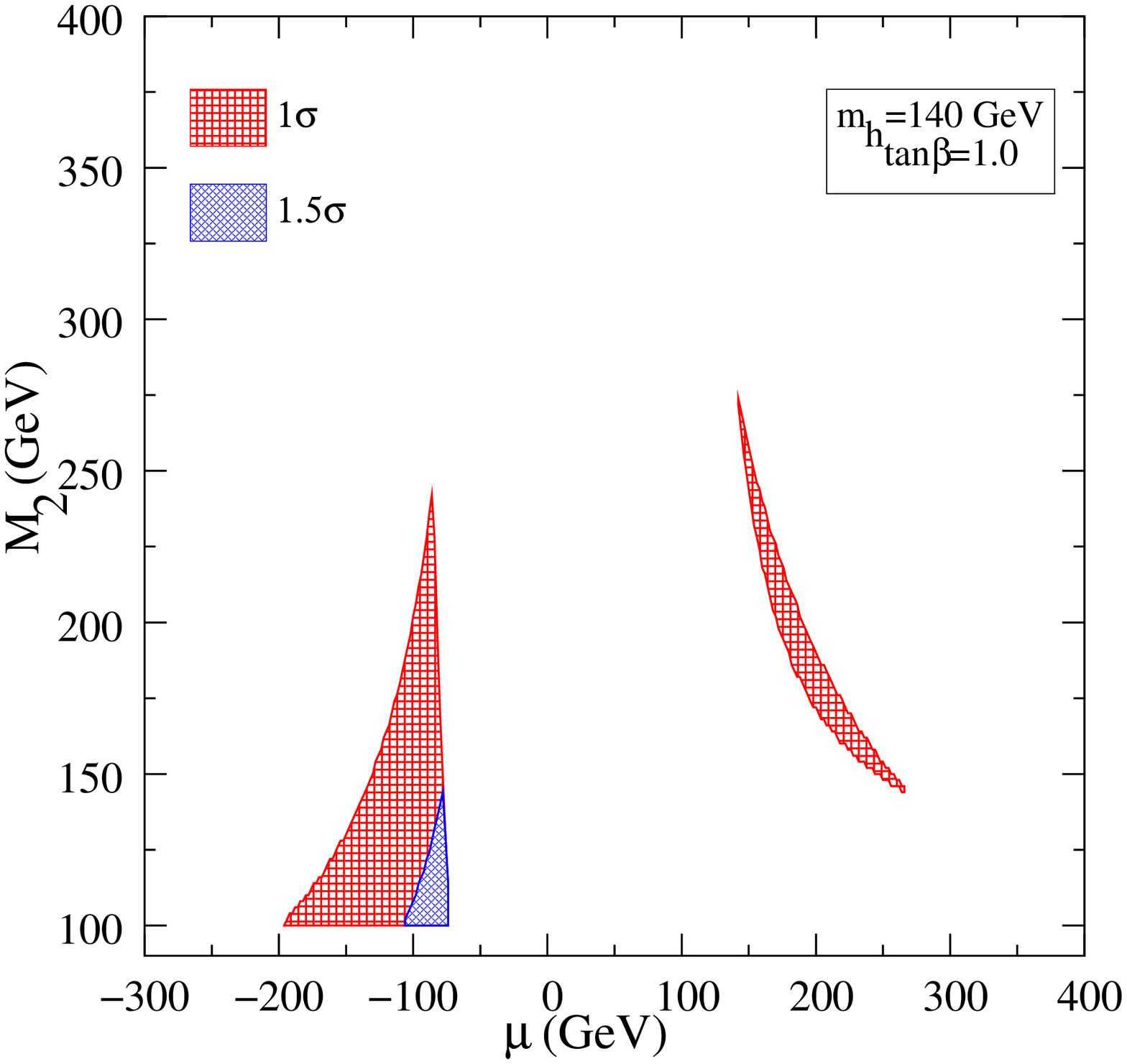,width=3in,height=3in}
\epsfig{file=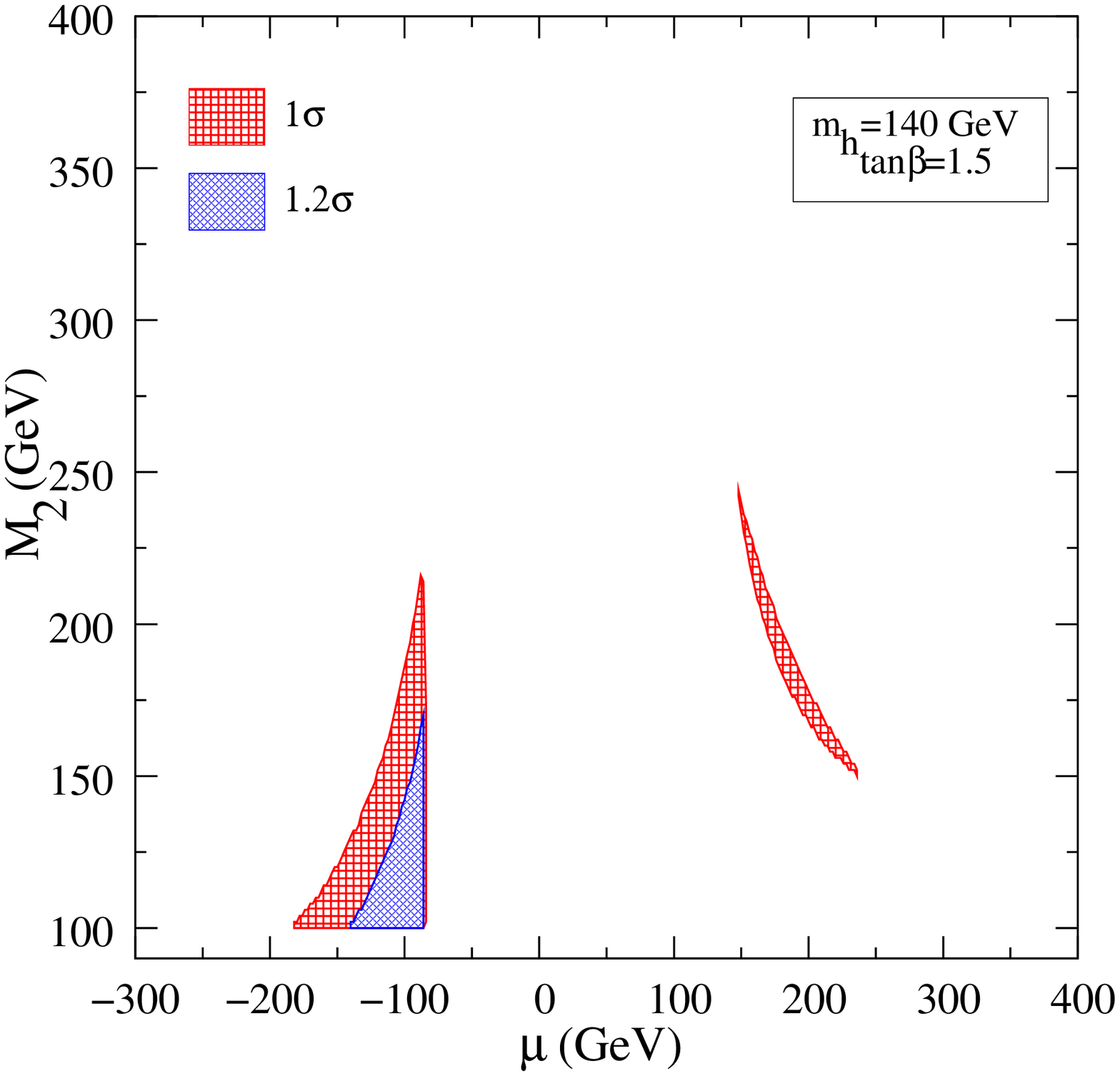,width=3in,height=3in}
\caption{\it{Contour plots for $m_h = 140$ GeV 
assuming PDF + scale uncertainty $=15\%$.}}
\label{15per140}
\end{center}
\end{figure}
%------------15 Percent-- and M_h=150 GeV---------------------
\begin{figure}
\begin{center}
\epsfig{file=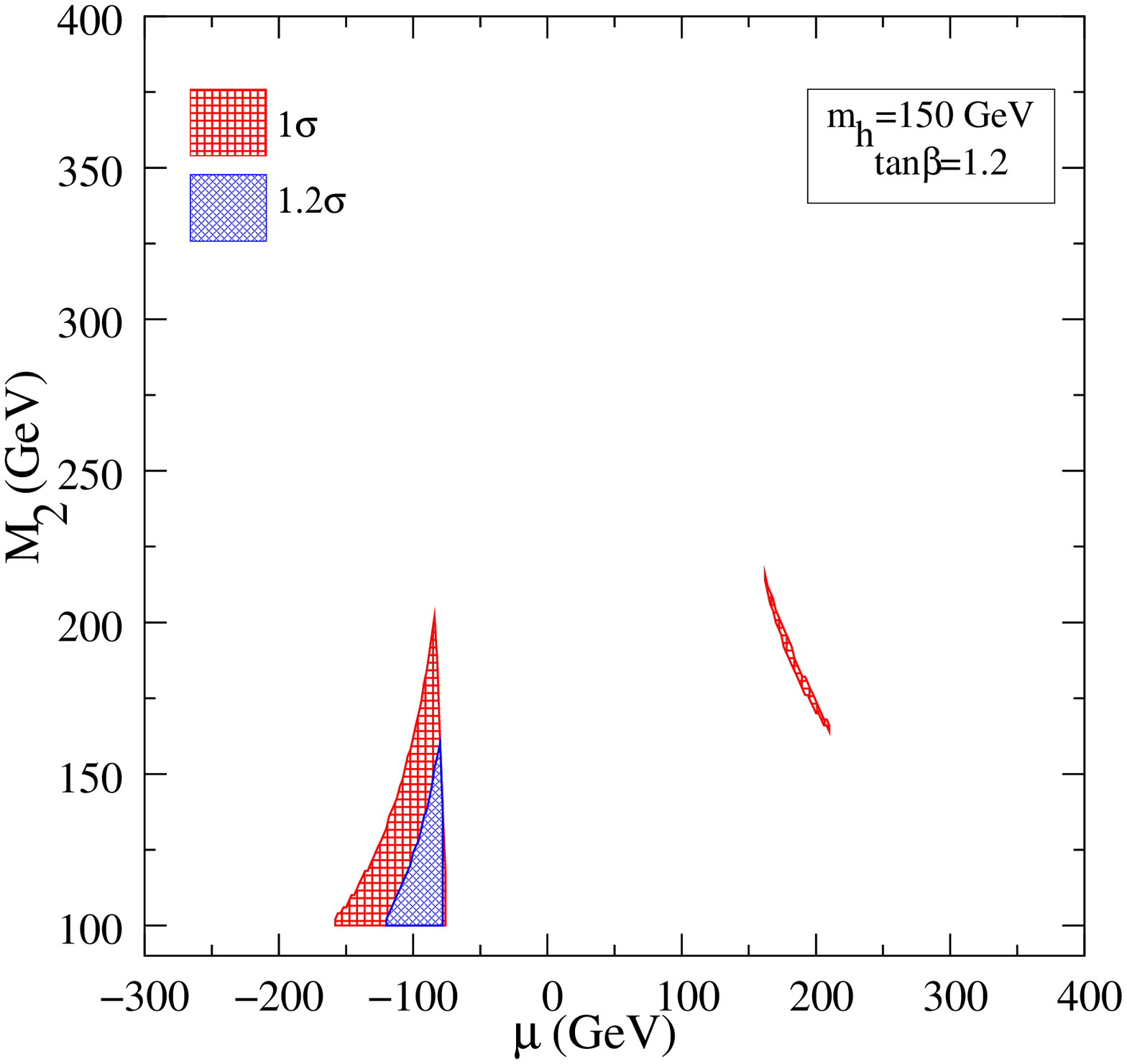,width=3in,height=3in}
\epsfig{file=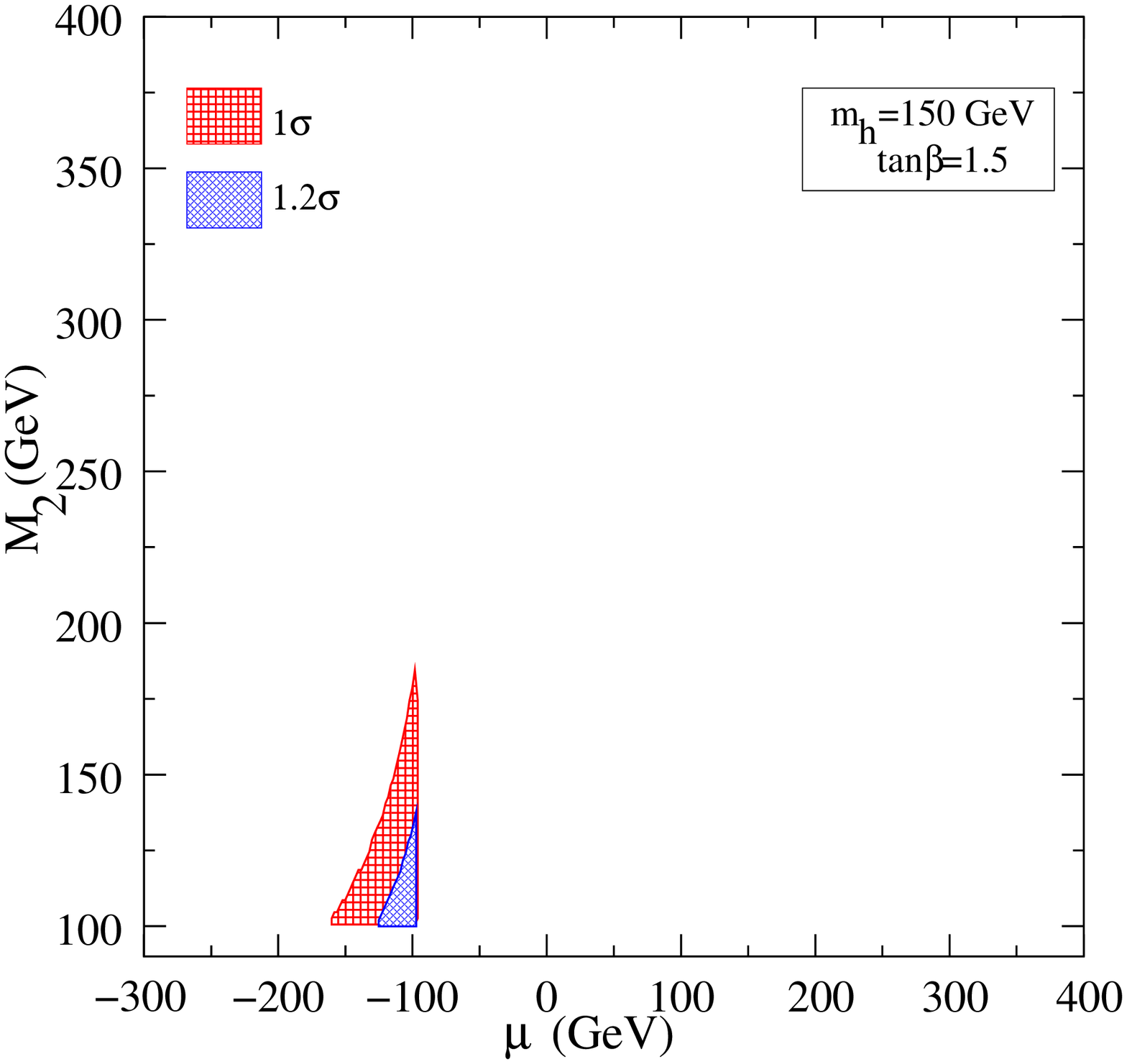,width=3in,height=3in}
\caption{\it{Contour plots for $m_h = 150$ GeV 
assuming PDF + scale uncertainty $=15\%$.}}
\label{15per150}
\end{center}
\end{figure}
%------------10 Percent-- and M_h=130 GeV---------------------
\begin{figure}
\begin{center}
\epsfig{file=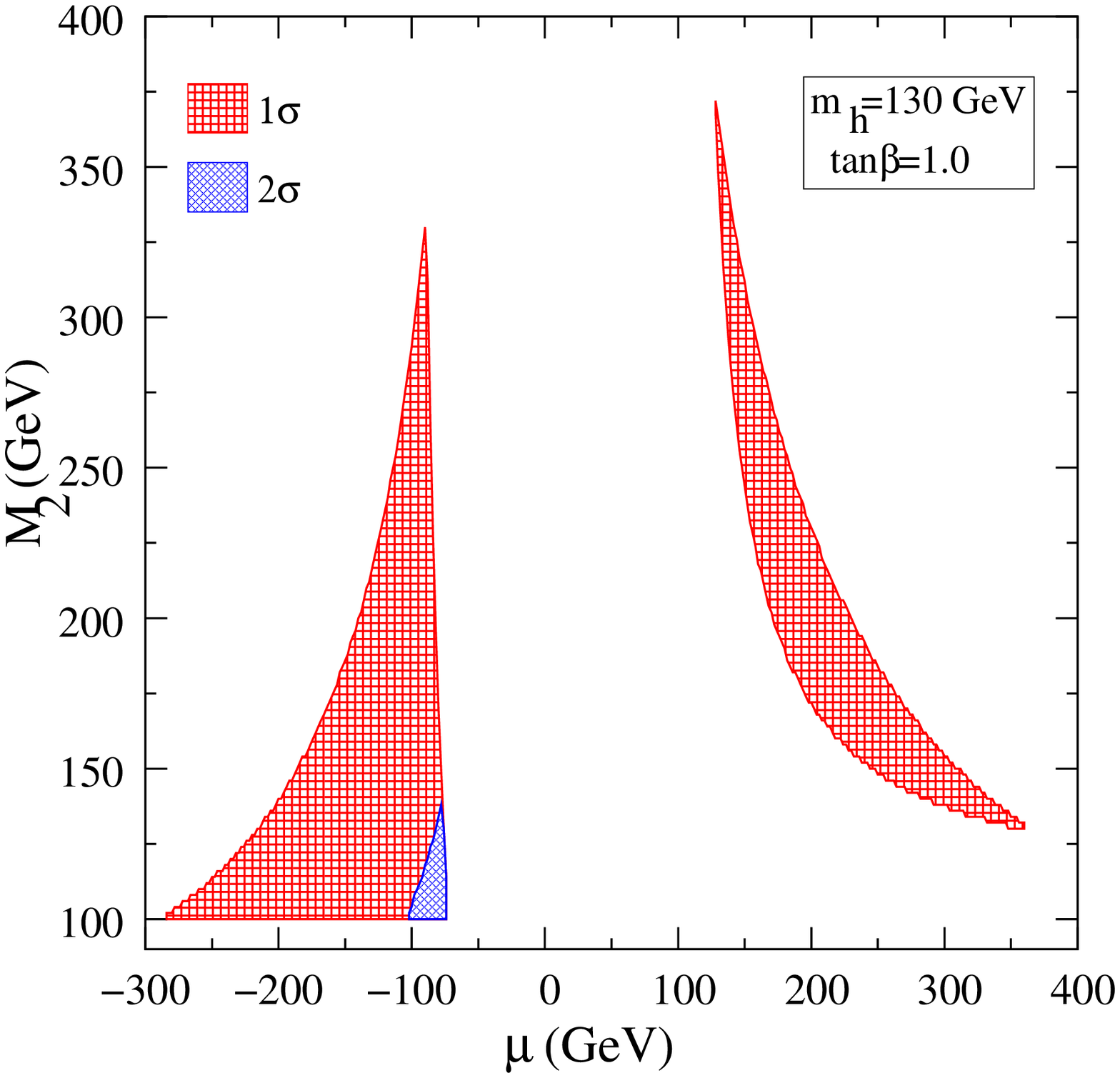,width=3in,height=3in}
\epsfig{file=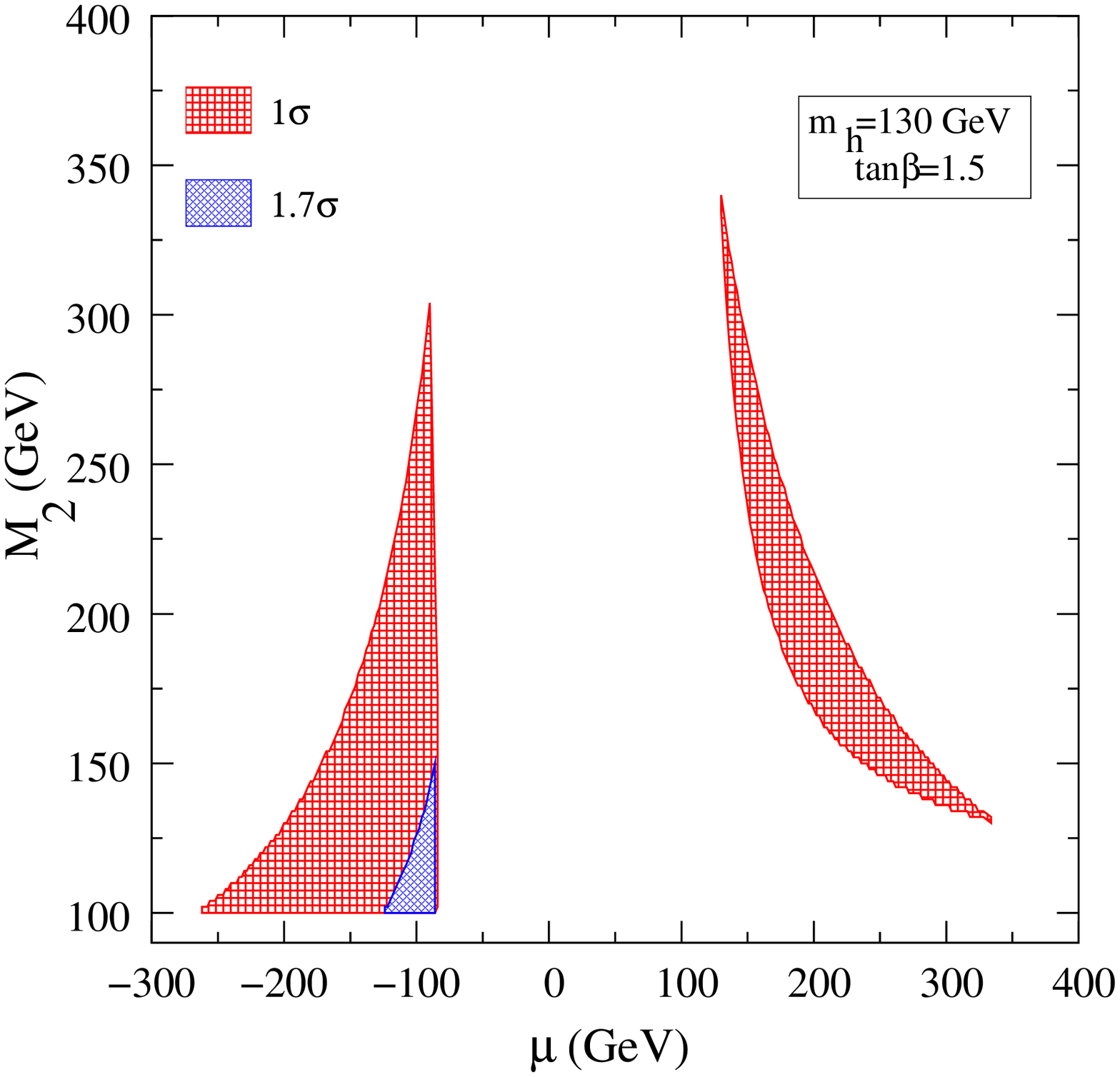,width=3in,height=3in}
\caption{\it{Contour plots for $m_h = 130$ GeV 
assuming PDF + scale uncertainty $=10\%$.}}
\label{10per130}
\end{center}
\end{figure}
%------------10 Percent-- and M_h=140 GeV---------------------
\begin{figure}
\begin{center}
\epsfig{file=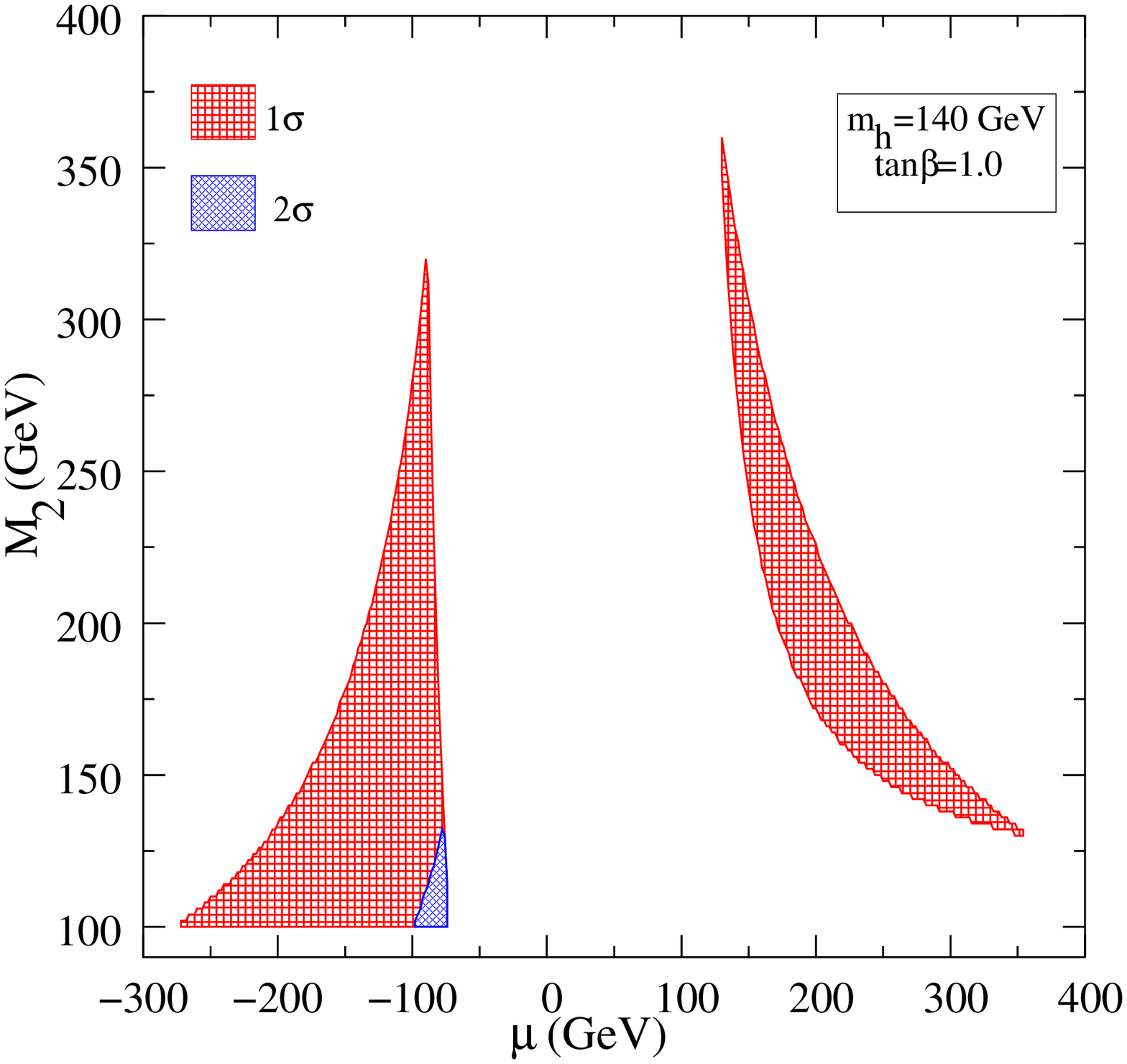,width=3in,height=3in}
\epsfig{file=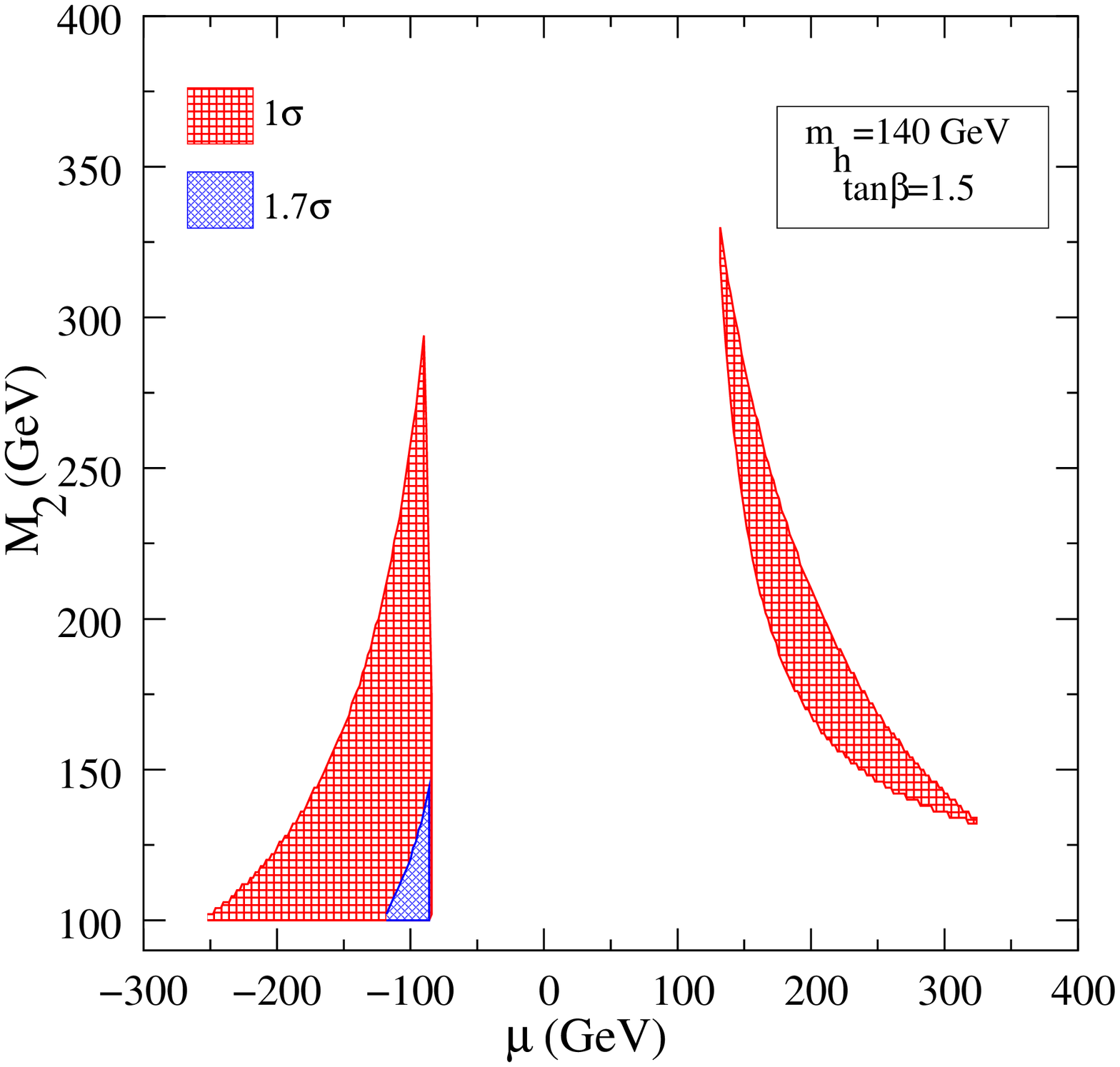,width=3in,height=3in}
\caption{\it{Contour plots for $m_h = 140$ GeV 
assuming PDF + scale uncertainty $=10\%$.}}
\label{10per140}
\end{center}
\end{figure}
%------------10 Percent-- and M_h=150 GeV---------------------
\begin{figure}
\begin{center}
\epsfig{file=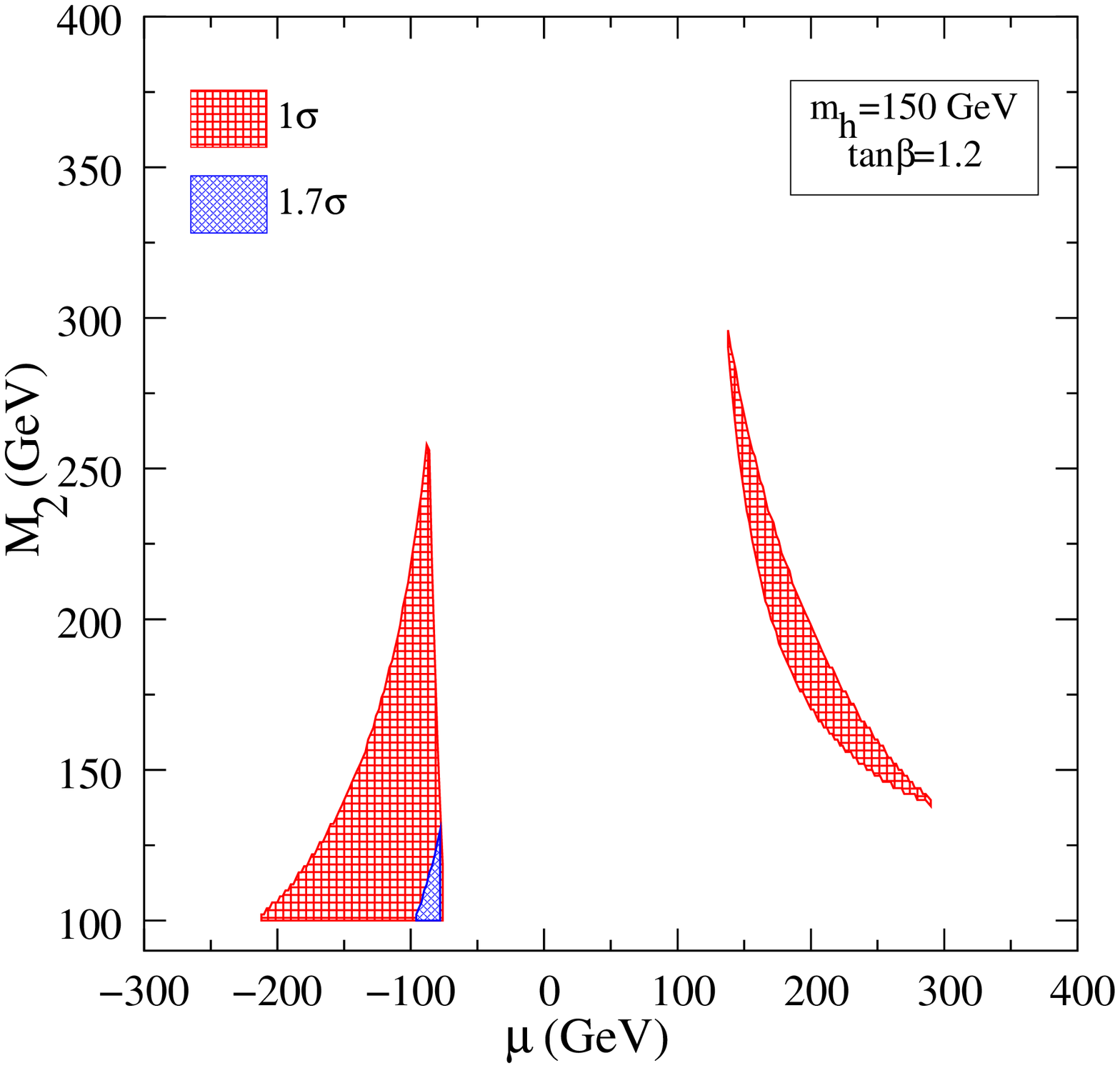,width=3in,height=3in}
\epsfig{file=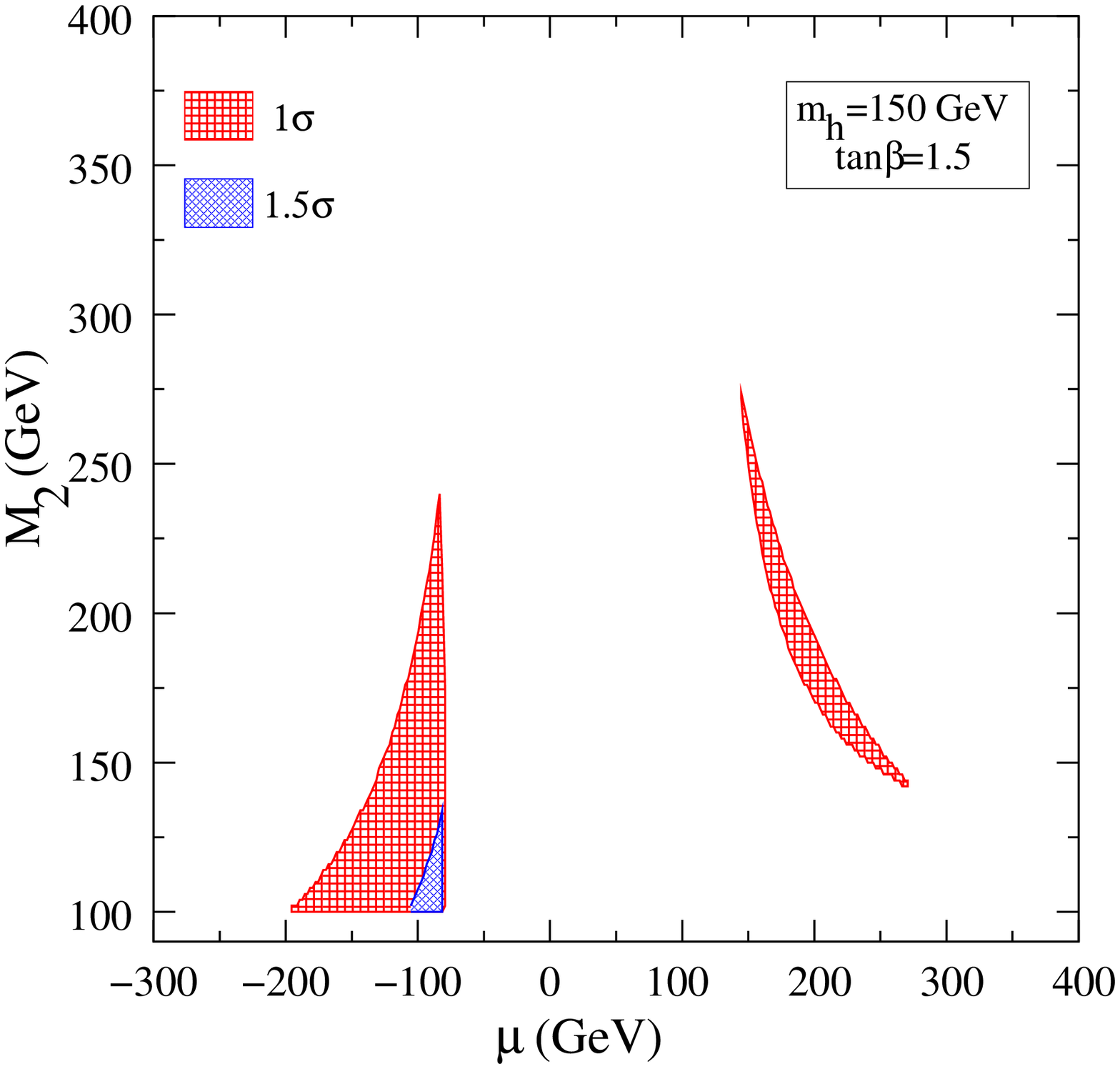,width=3in,height=3in}
\caption{\it{Contour plots for $m_h = 150$ GeV 
assuming PDF + scale uncertainty $=10\%$.}}
\label{10per150}
\end{center}
\end{figure}
%------------end---------------------------------------------
\end{document}